\documentclass{aa}

\usepackage{rotate,psfig}

\begin{document}

\thesaurus{03	    
              (11.01.2;			
               13.18.1;			
               11.17.3; 		
 )} 

\title{VLBI imaging of extremely high redshift quasars at 5~GHz}

\author{Z.~Paragi\inst{1,2}
\and
S.~Frey\inst{1}
\and
L.~I.~Gurvits\inst{2,3}
\and
K.~I.~Kellermann\inst{4}
\and
R.~T.~Schilizzi\inst{2,5}
\and 
R.~G.~McMahon\inst{6}
\and
I.~M.~Hook\inst{7}
\and
I.~I.~K.~Pauliny-Toth\inst{8}
}

\offprints{Z. Paragi, 1st address (paragi@sgo.fomi.hu)}

\institute{F\"OMI Satellite Geodetic Observatory, P.O. Box 546, H--1373
Budapest, Hungary 
\and
Joint Institute for VLBI in Europe, P.O. Box 2, NL--7990 AA
Dwingeloo, The Netherlands 
\and
Astro Space Center of P.N. Lebedev Physical Institute,
Moscow 117924, Russia
\and
National Radio Astronomy Observatory, 520 Edgemont Road,
Charlottesville, VA 22903-2475, USA 
\and
Leiden Observatory, P.O. Box 9513, NL--2300 RA Leiden, The Netherlands
\and
Institute of Astronomy, Madingley Road, Cambridge CB3 0HA, United
Kingdom
\and
European Southern Observatory, D--85748 Garching, Germany
\and 
Max-Planck-Institut f\"ur Radioastronomie, Auf dem H\"ugel 69, D--53121
Bonn, Germany
}

\date{Received 13 November 1998 / 
      Accepted 8 January 1999}

\authorrunning{Z. Paragi et al.}
\titlerunning{VLBI imaging of extremely high redshift quasars at 5~GHz}
\maketitle

\vspace{0.5cm}

%
%
%
%
%
%
%

\begin{abstract}

We present very long baseline interferometry (VLBI) images of ten 
very high redshift ($z>3$) quasars at
5~GHz.  The sources \object{0004+139}, \object{0830+101}, \object{0906+041}, 
\object{0938+119} and \object{1500+045} were observed in September 1992 
using a global VLBI array, while \object{0046+063}, \object{0243+181},
\object{1338+381}, \object{1428+423} and \object{1557+032} were observed in
October 1996 with the European VLBI Network and Hartebeesthoek, South
Africa.  Most of the sources are resolved and show asymmetric
structure.  The sample includes \object{1428+423}, the most distant radio loud
quasar known to date ($z=4.72$). It is barely resolved with an angular
resolution of about 2.0$\times$1.4 mas.


\end{abstract}

\section{Introduction}

There are about fifty known radio loud quasars at redshift $z > 3$ with
a total flux density at 5 GHz $S_5 \ga 100$ mJy. Some of them have been imaged 
at 5~GHz with VLBI (Gurvits et al. \cite{LIG92}, \cite{LIG94}; Xu et al. \cite{XUW95};
Taylor et al. \cite{TAY94}; Frey et al. \cite{FS97}; Udomprasert et al. \cite{UDO97}). 
Here we present first epoch VLBI images of a further ten $z > 3$ 
quasars. We show that their structural properties are similar
to those of other known sources at $z>3$. The present sample includes
the most distant radio loud quasar known to date, \object{1428+423} at $z=4.72$
(Hook \& McMahon \cite{H&M98}).

Our interest in studying the milliarcsecond radio structures in high
redshift quasars is motivated in part by their potential usefulness for
cosmological tests (e.g. Kellermann \cite{KIK93}; Gurvits et al. \cite{LIG99}).
Recent analysis of a sample of 151 quasars imaged at 5 GHz with
milliarcsecond resolution has led to the conclusion that a simple
assumption about the spectral properties of ``cores'' and ``jets'' can
explain the apparent greater compactness of the sources at higher
redshift (Frey et al. \cite{FS97}; Gurvits et al. \cite{LIG99}). 
However, this result is based on a sample
with considerable spread of structural properties on milliarcsecond
scale. More data on the milliarcsecond radio structures, especially
at high redshift are needed to study the structural properties of the
quasars as well as various cosmological models to test.

\section{Observations, calibration and data reduction}

Five sources (\object{0004+139}, \object{0830+101}, \object{0906+041}, 
\object{0938+119} and \object{1500+045}) were
observed during a single 24 hour observing run using a global VLBI
array on 27/28 September 1992. Another five sources (\object{0046+063},
\object{0243+181}, \object{1338+381}, \object{1428+423} and \object{1557+032}) 
were observed with the
European VLBI Network (EVN) and the Hartebeesthoek Radio Astronomical
Observatory 26~m antenna in South Africa on 25/26 and 27/28 October
1996. Source coordinates, redshifts and total flux densities at 6~cm
are given in Table~\ref{Sou}. The parameters of the radio telescopes used
in the two experiments are shown in Table~\ref{Sta}. The observations were made
at 5~GHz in left circular polarization. Data were recorded using the
Mk~III VLBI system in Mode B with 28~MHz total bandwidth, and
correlated at the MPIfR correlator in Bonn, Germany.

\begin{center}

\begin{table*}

\caption{Source parameters}
\label{Sou}
\begin{tabular}{lllllllll}
\hline
\noalign{\smallskip}
Source  & \multicolumn{3}{c}{RA (J2000)} & \multicolumn{3}{l}{Dec (J2000)}  & \,\,\,\,\,\,\,\,$S^{\rm a}$ & \,\,\,\,\,$z$ \\
         & (h)    & (m)      & (s)   & (\degr)  & (\arcmin) & (\arcsec) & \,\,\,\,(mJy)&  \\
\noalign{\smallskip}
\hline
\noalign{\smallskip}
0004+139 &  00    &  06      & 57.536 & 14       & 15        & 46.750    &  $152\pm14$ &  $3.25^{\rm b}$ \\ 
0046+063 &  00    &  48      & 58.740 & 06       & 40        & 05.900    &  $211\pm19$ &  $3.52^{\rm b}$ \\ 
0243+181 &  02    &  46      & 11.830 & 18       & 23        & 30.200    &  $220\pm20$ &  $3.59^{\rm b}$ \\ 
0830+101 &  08    &  33      & 22.514 & 09       & 59        & 41.140    & \,\,\,$93\pm9$ & $3.75^{\rm c}$ \\ 
0906+041 &  09    &  09      & 15.915 & 03       & 54        & 42.980    &  $111\pm11$ &  $3.20^{\rm d}$\\ 
0938+119 &  09    &  41      & 13.562 & 11       & 45        & 36.210    &  $123\pm12$ &  $3.177^{\rm e}$ \\ 
1338+381 &  13    &  40      & 22.952 & 37       & 54        & 43.833    &  $211\pm23$ &  $3.10^{\rm f}$ \\ 
1428+423 &  14    &  30      & 23.742 & 42       & 04        & 36.503    &  $337\pm30$ &  $4.72^{\rm g}$ \\ 
1500+045 &  15    &  03      & 28.886 & 04       & 19        & 48.980    &  $147\pm14$ &  $3.67^{\rm h}$ \\ 	 
1557+032 &  15    &  59      & 30.973 & 03       & 04        & 48.257    &  $414\pm37$ &  $3.90^{\rm h}$ \\ 
\noalign{\smallskip}
\hline
\end{tabular}

\begin{list}{}{}
\item[$^{\rm a}$] total flux density at 5~GHz from Gregory et al. \cite{GRE96} \
\item[$^{\rm b}$] Hook \cite{HOO94}; Hook \& McMahon (in prep.) \
\item[$^{\rm c}$] Oren \& Wolfe \cite{O&W95} \
\item[$^{\rm d}$] Brinkmann et al. \cite{BRI97} \
\item[$^{\rm e}$] Osmer et al. \cite{OSM94} \
\item[$^{\rm f}$] Hook et al. \cite{HOO95} \
\item[$^{\rm g}$] Hook \& McMahon \cite{H&M98} \
\item[$^{\rm h}$] McMahon et al. \cite{MCM94} \

\end{list}

\end{table*}

\end{center}

\begin{center}

\begin{table}

\caption{VLBI telescopes in the September 1992 (top) and October 1996 experiment (bottom) and their characteristics at 5~GHz}
\label{Sta}
\begin{tabular}{lcc}
\hline
\noalign{\smallskip}
Radio telescope	& Diameter (m) & SEFD$^{\rm a}$ (Jy)  \\ 
\noalign{\smallskip}
\hline
\noalign{\smallskip}
Effelsberg	   & 100     & 20	   \\
Medicina           & 32      & 296         \\
Onsala		   & 25	     & 780         \\
Westerbork & 93$^{\rm b}$    & 108         \\
NRAO Green Bank    & 43      & 77          \\
Haystack           & 37      & 533         \\
VLBA Owens Valley  & 25      & 289         \\
VLA        & $115^{\rm b}$   & 5.3         \\
\noalign{\smallskip}
\hline
\noalign{\smallskip}
Effelsberg	   & 100     & 20	   \\
Jodrell Bank Mk2   & 26      & 320         \\
Noto		   & 32	     & 260         \\
Torun              & 32      & 220         \\
Simeiz             & 22      & 3000        \\
Sheshan            & 25      & 520         \\
Nanshan            & 25      & 350         \\
Hartebeesthoek     & 26      & 790         \\
\noalign{\smallskip}
\hline
\end{tabular}

\begin{list}{}{}
\item[$^{\rm a}$] System Equivalent Flux Density \
\item[$^{\rm b}$] The telescope was used in phased array mode; \
\item[\,\,\,] an equivalent diameter is given.\\
\end{list}

\end{table}

\end{center}

Initial calibration was done using the NRAO AIPS package (Cotton \cite{COT95};
Diamond \cite{DIA95}). Clock offset and instrumental delay errors were
corrected using the strong sources \object{0804+499} and \object{0235+164} in the
global and the EVN experiments, respectively. Data were fringe-fitted using
AIPS using 5 minute solution intervals.  We used the system
temperatures measured during the observations and previously determined
gain curves for each telescope for the initial amplitude calibration,
which was then adjusted using amplitude calibrator sources, based on total
flux density values measured nearly contemporaneously to our observations 
with the Effelsberg telescope.
For the September 1992 experiment, this was also checked using VLA
data obtained in parallel with our VLBI observation.  Total flux
densities determined from VLBI images were typically 10-15\% smaller than
those determined from the VLA observations, which may indicate either
the presence of extended structures undetectable with VLBI or
residual calibration errors.

The Caltech DIFMAP program (Shepherd et al. \cite{SHE94}) was used for
self--calibration and imaging, starting with point source models with
flux densities consistent with the zero--spacing values. 
RMS image noises (3~$\sigma$) were 0.2-0.4 and 0.6-1.0~mJy/beam 
(depending on the telescopes' performance and integrated on--source time) 
for the global and the EVN experiments, respectively. Plots of
self--calibrated correlated flux densities as a function of projected
baseline length, as well as clean images resulting from the DIFMAP
imaging process are shown in Fig.~\ref{map} for both experiments. 
Image parameters are listed in Table~\ref{Mapparam}.
All sources but the most distant one, \object{1428+423}, appear to be well 
resolved and most of them show asymmetric structure.

We performed model fitting in DIFMAP using self--calibrated $uv$-data
in order to quantitatively compare these sources with other extremely
high redshift quasars.  The results of model fitting are listed in
Table~\ref{Mod}. In all cases we fixed the first component at the phase center.
While we searched for the simplest possible model (i.e. the smallest
possible number of Gaussian components), not all components can be
distinguished as separate features on the maps. In the case of
0004+139, we kept only one component for the extended emission because
the position angle of the beam lies close to the source structure
direction and the correlated flux density versus $uv$--distance plot
indicates the presence of a large component.

We also made $14\arcsec$ resolution VLA D configuration images of the
five sources observed in the 1992 global experiment. VLA data were
obtained at the same time as the phased array data used for the global
VLBI experiment. These images were made using the NRAO AIPS package
with typically 3--6 iterations of self--calibration and imaging. We
show VLA images of \object{0830+101} and \object{1500+045} in Fig.~\ref{VLA}a and \ref{VLA}b,
respectively. The other three sources appeared unresolved with the
VLA in our observations.

\begin{figure*}
\centerline{
\psfig{file=h1268map.f1a,width=6.5cm,bbllx=30pt,bblly=135pt,bburx=590pt,bbury=680pt,clip=}
\rotate[r]{
\psfig{file=h1268rad.f1a,width=6.0cm,bbllx=75pt,bblly=20pt,bburx=580pt,bbury=740pt,clip=}
}}
\end{figure*}

\begin{figure*}
\centerline{
\psfig{file=h1268map.f1b,width=6.5cm,bbllx=30pt,bblly=135pt,bburx=590pt,bbury=680pt,clip=}
\rotate[r]{
\psfig{file=h1268rad.f1b,width=6.0cm,bbllx=75pt,bblly=20pt,bburx=580pt,bbury=740pt,clip=}
}}
\end{figure*}

\begin{figure*}
\centerline{
\psfig{file=h1268map.f1c,width=6.5cm,bbllx=30pt,bblly=135pt,bburx=590pt,bbury=680pt,clip=}
\rotate[r]{
\psfig{file=h1268rad.f1c,width=6.0cm,bbllx=75pt,bblly=20pt,bburx=580pt,bbury=740pt,clip=}
}}

\caption{{\it (continued on the next page)} }
\end{figure*}

\addtocounter{figure}{-1}

\begin{figure*}
\centerline{
\psfig{file=h1268map.f1d,width=6.5cm,bbllx=30pt,bblly=135pt,bburx=590pt,bbury=680pt,clip=}
\rotate[r]{
\psfig{file=h1268rad.f1d,width=6.0cm,bbllx=75pt,bblly=20pt,bburx=580pt,bbury=740pt,clip=}
}}
\end{figure*}

\begin{figure*}
\centerline{
\psfig{file=h1268map.f1e,width=6.5cm,bbllx=30pt,bblly=135pt,bburx=590pt,bbury=680pt,clip=}
\rotate[r]{
\psfig{file=h1268rad.f1e,width=6.0cm,bbllx=75pt,bblly=20pt,bburx=580pt,bbury=740pt,clip=}
}}
\end{figure*}
 
\begin{figure*}
\centerline{
\psfig{file=h1268map.f1f,width=6.5cm,bbllx=30pt,bblly=135pt,bburx=590pt,bbury=680pt,clip=}
\rotate[r]{
\psfig{file=h1268rad.f1f,width=6.0cm,bbllx=75pt,bblly=20pt,bburx=580pt,bbury=740pt,clip=}
}}

\caption{{\it (continued on the next page)} }
\end{figure*}

\addtocounter{figure}{-1}

\begin{figure*}
\centerline{
\psfig{file=h1268map.f1g,width=6.5cm,bbllx=30pt,bblly=135pt,bburx=590pt,bbury=680pt,clip=}
\rotate[r]{
\psfig{file=h1268rad.f1g,width=6.0cm,bbllx=75pt,bblly=20pt,bburx=580pt,bbury=740pt,clip=}
}}
\end{figure*}
 
\begin{figure*}
\centerline{
\psfig{file=h1268map.f1h,width=6.5cm,bbllx=30pt,bblly=135pt,bburx=590pt,bbury=680pt,clip=}
\rotate[r]{
\psfig{file=h1268rad.f1h,width=6.0cm,bbllx=75pt,bblly=20pt,bburx=580pt,bbury=740pt,clip=}
}}
\end{figure*}

\begin{figure*}
\centerline{
\psfig{file=h1268map.f1i,width=6.5cm,bbllx=30pt,bblly=135pt,bburx=590pt,bbury=680pt,clip=}
\rotate[r]{
\psfig{file=h1268rad.f1i,width=6.0cm,bbllx=75pt,bblly=20pt,bburx=580pt,bbury=740pt,clip=}
}}

\caption{{\it (continued on the next page)} }
\end{figure*}

\addtocounter{figure}{-1}

\begin{figure*}
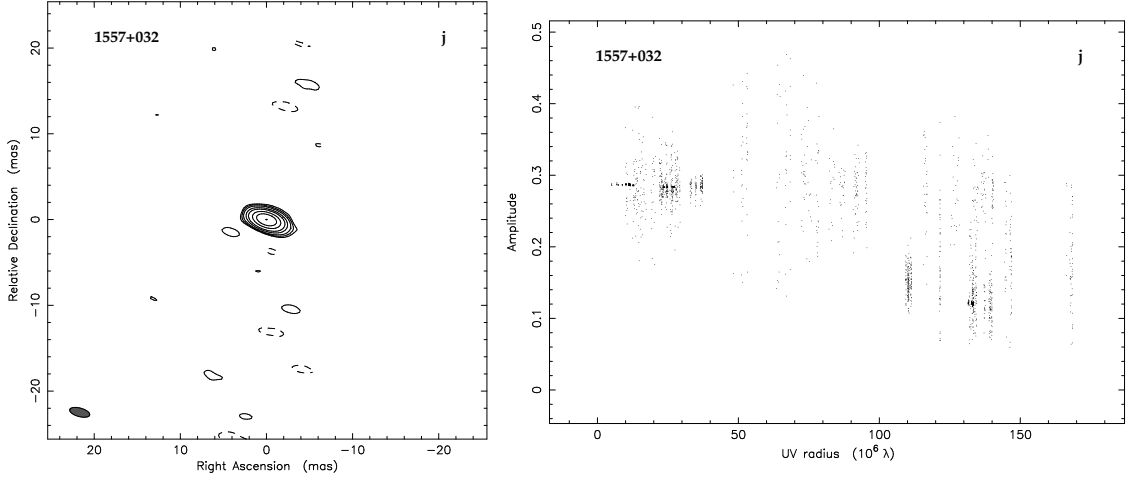

\centerline{
\psfig{file=h1268map.f1j,width=6.5cm,bbllx=30pt,bblly=135pt,bburx=590pt,bbury=680pt,clip=}
\rotate[r]{
\psfig{file=h1268rad.f1j,width=6.0cm,bbllx=75pt,bblly=20pt,bburx=580pt,bbury=740pt,clip=}
}}

\caption{{\it (continued)} ``Naturally'' weighted 5~GHz images (left)
and correlated flux density (Jy) versus projected baseline length
(right) for {\bf a} 0004+139, {\bf b} 0046+063, {\bf c} 0243+181,
{\bf d} 0830+101, {\bf e} 0906+041, {\bf f} 0938+119, {\bf g} 1338+381,
{\bf h} 1428+423, {\bf i} 1500+045 and {\bf j} 1557+032. Map parameters are given in Table 3}
\label{map}
\end{figure*}

\begin{table*}

\caption{Parameters of VLBI maps in Fig. 1a-j}
\label{Mapparam}
\begin{tabular}{llcrrrc}
\hline
\noalign{\smallskip}
Source   & Contour levels                              & Peak brightness & \multicolumn{3}{c}{Restoring beam} & Observing epoch \\
         &                                             &                 & $\theta_{max}$ & $\theta_{min}$ & PA      &       \\
         & (\% of the peak brightness)                 & (mJy/beam)      & (mas)       & (mas)      & (\degr) &       \\
\noalign{\smallskip}
\hline
\noalign{\smallskip}
0004+139 & $-$0.25, 0.25, 0.5, 1, 2, 5, 10, 25, 50, 99 &  \,\,50             & 7.8         & 0.8        & $-$13   & Sep 1992 \\
0046+063 & $-$0.6, 0.6, 1.2, 2.5, 5, 10, 25, 50, 99    &  104            & 2.6         & 1.2        & 74      & Oct 1996 \\ 
0243+181 & $-$0.4, 0.4, 0.8, 2, 5, 10, 25, 50, 99      &  119            & 1.9         & 1.4        & 66      & Oct 1996 \\  
0830+101 & $-$0.5, 0.5, 1, 2, 5, 10, 25, 50, 99        &  \,\,74             & 8.2         & 0.8        & $-$11   & Sep 1992 \\  
0906+041 & $-$0.4, 0.4, 0.8, 2, 5, 10, 25, 50, 99      &  \,\,44             & 11.8        & 0.8        & $-$10   & Sep 1992 \\  
0938+119 & $-$0.6, 0.6, 1.2, 2.5, 5, 10, 25, 50, 99    &  \,\,49             & 8.7         & 1.2        & $-$17   & Sep 1992 \\  
1338+381 & $-$0.6, 0.6, 1.2, 2.5, 5, 10, 25, 50, 99    &  148            & 2.0         & 1.5        & 81      & Oct 1996 \\  
1428+423 & $-$0.5, 0.5, 1, 2, 5, 10, 25, 50, 99        &  161            & 2.0         & 1.4        & 38      & Oct 1996 \\ 
1500+045 & $-$0.25, 0.25, 0.5, 1, 2, 5, 10, 25, 50, 99 &  131            & 10.5        & 1.0        & $-$12   & Sep 1992 \\  
1557+032 & $-$0.5, 0.5, 1, 2, 5, 10, 25, 50, 99        &  228            & 2.4         & 1.0        & 75      & Oct 1996 \\
\noalign{\smallskip}
\hline
\end{tabular}
\begin{list}{}{}
\item[\,\,\,]   Note: $\theta_{max}$, $\theta_{min}$ and PA are restoring beam major axis, minor axis and position angle, respectively.\
\end{list}
\end{table*}


\section{Comments on individual sources}

\paragraph{\object{0004+139}}

The spectral indices of the source ($S\propto\nu^{\alpha}$ throughout this
paper) are $\alpha_{0.365}^{1.4}=-0.6$ and $\alpha_{1.4}^{4.85}=-0.4$
(White \& Becker \cite{W&B92}). It is unresolved with the VLA A-array
($\sim400$~mas resolution) at 5~GHz (Lawrence et al. \cite{LAW86}). 

Our VLBI image shows structure extending up to about 10~mas from the core to
the SE direction (Fig.~\ref{map}a). The position angle of the beam is not well suited to
resolve the fine details of this jet--like extension. The source is
unresolved with the VLA D-array in our experiment with $14\arcsec$
resolution.

\paragraph{\object{0046+063}}

The source has a flat radio spectrum between 1.4 and 4.85~GHz
($\alpha_{1.4}^{4.85}=-0.0$, White \& Becker \cite{W&B92}). Our VLBI image
shows a dominant central component and a prominent secondary component
separated by 3.8~mas from the core in the NE direction (Fig.~\ref{map}b).

\paragraph{\object{0243+181}}

The spectral index of this quasar is $\alpha_{1.4}^{4.85}=0.1$ (White
\& Becker \cite{W&B92}). Apart from the compact core, there is a weak extended
feature 4.9~mas to the South (Fig.~\ref{map}c).

\paragraph{\object{0830+101}}

\begin{figure*}
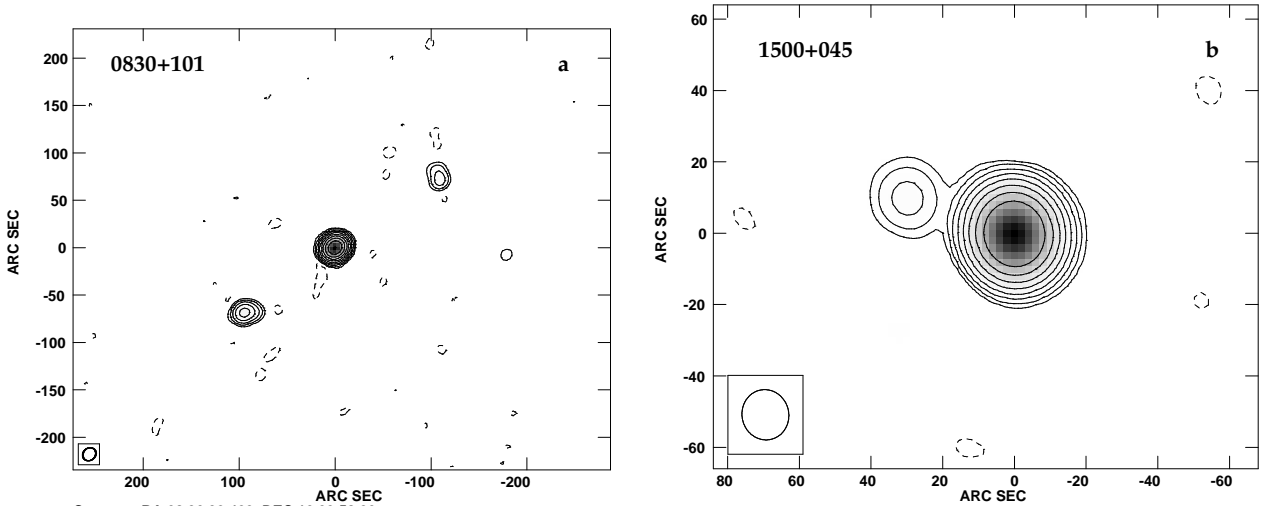

\centerline{
\psfig{file=h1268.f2a,width=8.5cm,bbllx=30pt,bblly=197pt,bburx=590pt,bbury=622pt,clip=}
\psfig{file=h1268.f2b,width=8.5cm,bbllx=30pt,bblly=188pt,bburx=590pt,bbury=632pt,clip=}
}
\caption{5~GHz VLA images of {\bf a} 0830+101 and {\bf b} 1500+045.
The contour levels are {\bf a} $-$1, 1, 2, 4, 8, 16, 32, 64, 128, 256,
512, 1024 $\times$ 0.1 mJy/beam and {\bf b} $-$1, 1, 2, 4, 8, 16, 32, 64,
128, 256 $\times$ 0.2 mJy/beam. The restoring circular beam is $14\arcsec$ (HPBW)}
\label{VLA}
\end{figure*}

The source is reported to be unresolved at 5~GHz with the VLA B--array
($\sim1.2\arcsec$ resolution), no extended emission has been found
within about $51\arcsec$ from the core (Lawrence et al. \cite{LAW86}). The
spectral index of the source is $\alpha_{1.4}^{4.85}=-0.3$ (White \&
Becker \cite{W&B92}). On the VLBI scale, it has two bright components near the core that perhaps
delineate a slightly curved jet extending up to $\sim$15~mas (Fig.~\ref{map}d).  Our VLA D-array
map shows two faint components about $2\arcmin$ from the core to the SE
and NW which resembles a classical double lobe structure (Fig.~\ref{VLA}a).
However, it is not clear from our VLA image whether these sources are
physically related to 0830+101 or they are chance coincidences.
The latter seems to be unlikely, but could not be ruled out based on our data.

\paragraph{\object{0906+041}}

Spectral indices of $\alpha_{0.365}^{1.4}=0.1$ and
$\alpha_{1.4}^{4.85}=-0.4$ are given by White \& Becker (\cite{W&B92}). If the
flux density of the source did not change between the epochs of
measurements this indicates that the source may be a Gigahertz Peaked
Spectrum (GPS) quasar. This object has been identified as a ROSAT X--ray
source (RXJ0909.2+0354). Its flux in the 0.1--2.4 keV range is
$f_{x}=9.9\pm2.7$~$10^{-13}$ ~erg~cm$^{-2}$~s$^{-1}$ (Brinkmann et al.
\cite{BRI95}). The source is unresolved with the VLA D-array at 5~GHz. On VLBI
scales, the core of 0906+041 is resolved with an extension to
the NE (Fig.~\ref{map}e). A secondary compact component is separated by about 10~mas from
the core.

\paragraph{\object{0938+119}}

This source is identified as a quasar by Beaver et al. (\cite{BEA76}) and has a
very steep optical continuum more typical of BL Lac objects (Baldwin et
al. \cite{BAL76}). The radio continuum peaks near 1~GHz
($\alpha_{0.365}^{1.4}=-0.0$ and $\alpha_{1.4}^{4.85}=-0.7$, White \&
Becker \cite{W&B92}). Neff \& Hutchings (\cite{N&H90}) found radio emission with the
VLA at 1.4~GHz on both sides of the radio core extending to $5\arcsec$
and $2\arcsec$ from the centre.  The source was studied in high energy
bands, however, only upper limits are available for X--ray and
$\gamma$--ray luminosities (Zamorani et al. \cite{ZAM81}; Fichtel et al. \cite{FIC94}).
The source is resolved by our observations and shows an extension of
about 5~mas to the East (Fig.~\ref{map}f). It is unresolved with the VLA in our
experiment.

\paragraph{\object{1338+381}}

This flat spectrum source ($\alpha_{1.4}^{4.85}=-0.0$, White \& Becker
\cite{W&B92}) is a candidate IERS radio reference frame object and serves as a
link to the HIPPARCOS stellar reference frame (Ma et al. \cite{MA97}).  It is
being monitored by geodetic VLBI networks at 2.3 and 8.4~GHz.  In our
5~GHz imaging experiment the source appears to be resolved and shows a
double structure elongated in the S-SW direction with the angular
separation of 3.65~mas (Fig.~\ref{map}g). The component position angle and separation are
in very good agreement with a recent 8.4~GHz global VLBI image by Bouchy et
al.  (\cite{BOU98}). Due to the lower resolution of our image we can not decide
whether their component ``c'' is present between the two dominant
components seen in our image.

\paragraph{\object{1428+423}}

The radio spectral indices of the quasar 1428+423 -- also known as
\object{GB1428+4217} (Fabian et al. \cite{FAB97}; Hook \& McMahon \cite{H&M98})
and B3~1428+422 (V\'{e}ron-Cetty and V\'{e}ron \cite{VER98}) -- are
$\alpha_{0.365}^{1.4}=0.5$ and $\alpha_{1.4}^{4.85}=-0.4$ (White \&
Becker \cite{W&B92}) which are typical for GPS sources. It is the third highest
redshift quasar known to date (Hook \& McMahon \cite{H&M98}, $z=4.72$) and the
most distant known radio loud quasar. The quasar was detected in X-rays
with the ROSAT High Resolution Imager in the (observed) 0.1--2.4~keV
band (Fabian et al. \cite{FAB97}) and with various ASCA detectors in the
(observed) band of 0.5--10~keV (Fabian et al. \cite{FAB98}). Both observations
are in agreement and indicate that the SED of this source is strongly
dominated by X- and $\gamma$-ray emission. The X-ray spectrum
is remarkably flat. The quasar might be the most luminous steady source
in the Universe, with an apparent luminosity in excess of $10^{47}$ erg s$^{-1}$.
The extreme X-ray luminosity of the quasar 1428+423 suggests
that the emission is highly beamed toward us (Fabian et al.
\cite{FAB97}, \cite{FAB98}).

Our VLBI image (Fig.~\ref{map}h) is in qualitative agreement with the relativistic
beaming model of the source. The quasar appears to be almost unresolved
with the VLA at 5~GHz (Laurent-Muehleisen et al. \cite{LAU97}) as well as by
our VLBI observations up to 170 M$\lambda$, which corresponds to an
angular resolution of 2.0$\times$1.4 mas. The other two $z > 4$ quasars
imaged with VLBI also appear unresolved (\object{1251$-$407} and
\object{1508+572}, Frey et al. \cite{FS97}), which suggests that the high $z$ quasars may
be systematically more compact then their less distant counterparts.
Alternatively, as suggested by Fabian et al. (\cite{FAB97}), the highly beamed
emission might be responsible for a selection effect resulting in
detection of an otherwise weaker population of extremely high redshift
quasars.

\paragraph{\object{1500+045}}
 
This source was detected as a $5\pm2.4$~mJy source at 240~GHz by
McMahon et al. (\cite{MCM94}) corresponding to $\alpha_{5}^{240}=-0.9$. The
spectral index between 1.4 and 4.85~GHz is
$\alpha_{1.4}^{4.85}=0.2$ (White \& Becker \cite{W&B92}). The source is
unresolved with the VLA B--array ($\sim1.2\arcsec$ resolution, Lawrence
et al. \cite{LAW86}).

Although the source is resolved, our VLBI image does not show any
structure (Fig.~\ref{map}i).  Our VLA image shows an extension to E-NE at about $33\arcsec$
(Fig.~\ref{VLA}b).

\paragraph{\object{1557+032}}

This quasar is an IERS Celestial Reference Frame candidate source (Ma
et al. \cite{MA97}).  It was also detected with the Parkes--Tidbinbilla
interferometer at 2.3~GHz (Duncan et al. \cite{DUN93}) and found to be compact
with a total flux density of 376~mJy. Our VLBI observations show that
the source is resolved but featureless (Fig.~\ref{map}j). There is no extended feature  found
down to 0.5\% of the peak brightness.

\begin{table*}
\caption{Fitted elliptical Gaussian model parameters of the source structures}
\label{Mod}
\begin{tabular}{lcrrrrrrrr}
\hline
\noalign{\smallskip}
Source     & Component & $S$    & $r$    & $\Theta$  & $a$\,\,\,\,  & $b/a$ & $\Phi$     & Agreement & $S_{j}/S_{c}$ \\
           &           & (mJy)  & (mas)  & (\degr)   & (mas) &       & (\degr)    & factor    &               \\
           &           & [1]    & [2]    & [3]       & [4]   & [5]   & [6]        & [7]       & [8]           \\
\noalign{\smallskip}
\hline
\noalign{\smallskip}
0004+139   & A         & 56       & 0.0     & --        & 2.5   & 0.0   & $-$32   & 0.94      & $-$           \\ 
           & B         & 37       & 4.0     & 156       & 5.8   & 3.6   & $-$28   &           &               \\
0046+063   & A         & 117      & 0.0     & --        & 0.7   & 0.3   & 17      & 1.06      & 0.40          \\ 
           & B         & 47       & 3.8     & 37        & 1.5   & 0.1   & 36      &           &               \\
0243+181   & A         & 96       & 0.0     & --        & 0.8   & 0.6   & 3       & 1.03      & 0.05          \\ 
           & B         & 48       & 0.4     & 5         & 1.5   & 0.0   & 1       &           &               \\
           & C         & 5        & 4.9     & 176       & 0.8   & 0.7   & 59      &           &               \\
0830+102   & A         & 81       & 0.0     & --        & 1.2   & 0.3   & 0       & 0.99      & 0.22          \\
           & B         & 18       & 4.0     & 120       & 2.8   & 0.0   & $-$33   &           &               \\
           & C         & 7        & 15.8    & 123       & 4.7   & 0.0   & $-$26   &           &               \\
0906+041   & A         & 35       & 0.0     & --        & 2.5   & 0.1   & $-$10   & 0.90      & 0.63          \\
           & B         & 22       & 2.1     & 4         & 1.5   & 0.0   & $-$70   &           &               \\
           & C         & 2        & 9.3     & 8         & 5.8   & 0.1   & $-$13   &           &               \\
0938+119   & A         & 50       & 0.0     & --        & 1.2   & 0.1   & 55      & 0.85      & $<$0.005       \\
           & B         & 40       & 0.6     & 106       & 4.1   & 0.3   & 51      &           &               \\
1338+381   & A         & 187      & 0.0     & --        & 1.1   & 0.6   & $-$10   & 1.04      & 0.32          \\ 
           & B         & 10       & 1.6     & 116       & 3.7   & 0.2   & $-$80   &           &               \\
           & C         & 60       & 3.7     & $-$163    & 1.2   & 0.3   & 13      &           &               \\
1428+423   & A         & 173      & 0.0     & --        & 0.6   & 0.5   & 11      & 0.95      & $<$0.005       \\ 
1500+045   & A         & 138      & 0.0     & --        & 0.7   & 0.0   & 28      & 0.95      & $-$           \\
           & B         & 11       & 0.7     & 97        & 3.3   & 0.2   & $-$11   &           &               \\
1557+032   & A         & 276      & 0.0     & --        & 0.7   & 0.0   & 2       & 1.11      & 0.05          \\ 
           & B         & 15       & 1.1     & $-$154    & 4.6   & 0.0   & 76      &           &               \\
\noalign{\smallskip}
\hline
\end{tabular}
\begin{list}{}{}
\item[\,\,\,]   [1] $S$ flux density,\
\item[\,\,\,]   [2] $r$ angular separation from the central component,\
\item[\,\,\,]   [3] $\Theta$ position angle (Position angles are measured from the North through East),\
\item[\,\,\,]   [4] $a$ component major axes,\
\item[\,\,\,]   [5] $b/a$ ratio of the component minor and major axes,\
\item[\,\,\,]   [6] $\Phi$ component major axis position angle,\
\item[\,\,\,]   [7] The agreement factor is the square root of reduced $\chi^{2}$, see e.g. Pearson (\cite{PEA95}),\
\item[\,\,\,]   [8] $S_{j}/S_{c}$ jet to core flux density ratio.
\end{list}
\end{table*}

\section{Discussion}

Frey et al. (\cite{FS97}) studied the parsec scale structural properties of
radio loud QSO's using a sample of 151 quasars in the redshift range of
$0.2<z<4.5$ observed with sufficiently high resolution at 5~GHz.
They determined the flux density ratios of the brightest ``jet'' and ``core''
components ($S_j$/$S_c$) of the sources. The typical angular resolution
of those VLBI observations was $\sim$1~mas. Because the linear resolution is better
for the lowest redshift sources, they introduced a linear size limit to 
distinguish between jet and core components in order to compare the same
linear sizes at different redshifts. One milliarcsecond sets the linear
resolution to 7~pc for $z\ga1$ sources up to the highest redshifts represented
in the sample ($H_{0}$=80~km~s$^{-1}$\,Mpc$^{-1}$ and $q_{0}$=0.1 were used 
to calculate linear sizes; the angular size of a fixed linear size is practically
constant at $z\ga1$ for plausible cosmological models). The value of 7~pc was not used in 
any quantitative way in their analysis, just as a threshold between cores and jets.
Only components outside the core region were
considered as jet components. They found a weak overall trend of a
decreasing jet to core flux density ratio with increasing redshift
which may be explained by the combined effect of the shifts of the
emitting frequencies at different redshifts compared to the 5~GHz
observing frequency and the different characteristic spectral indices
in cores and jets.

We followed Frey et al. (\cite{FS97}) and calculated the jet to core flux
density ratios ($S_j$/$S_c$) for the $z>3$ sources presented in this
paper (last column of Table~4). We had to exclude two sources  from the
analysis, \object{0004+139} and \object{1500+045}. In the former case, the angular
resolution in the direction of the expected jet structure is
considerably greater than 1~mas. The quasar \object{1500+045} may also have jet
structure which is not observable in our data due to the unfortunate
orientation of the 10.5 mas restoring beam. The $S_j$/$S_c$ values for
the other three sources observed in the 1992 global experiment
(\object{0830+101}, \object{0906+041} and \object{0938+119}) should also be interpreted with
caution since the restoring beam is very elongated. However, we derived
tentative $S_j$/$S_c$ values because the direction of the jet structure
indicated by our maps are nearly perpendicular to the major axis of the
beam and the resolution in this direction is about 1~mas. In the case
of \object{0938+119} and \object{1428+423}, an upper limit of the jet flux density was
calculated based on the beam sizes and the 3$\sigma$ RMS noises on our images.

\begin{figure}
\centerline{
\rotate[r]{
\psfig{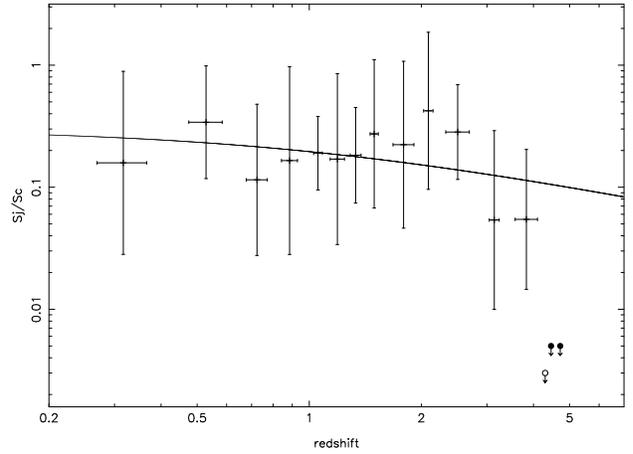}
}}
\caption{Median jet to core flux density ratios versus redshift for 159
quasars of Frey et al. (1997) and this paper. Values are grouped into
13 nearly equally populated bins (12-13 sources per bin). 
The solid curve represents the best fit to the 13 median values. 
Circles indicate the upper limits of $S_j/S_c$
for the three $z>4$ quasars imaged with VLBI at 5~GHz to date (see Sect.~4.)}
\label{SjSc}
\end{figure}

We added our eight new sources to the sample of Frey et al. (\cite{FS97}). 
The median $S_j/S_c$ values as function of redshift are
shown in Fig.~\ref{SjSc}.  The data for all 159 sources are evenly grouped into 13 bins.
Error bars indicate the mean absolute deviation of data points from the
median within each bin. Upper limits and measured values are treated
similarly. However, the plotted error bars are indicative of the scatter
of the data.
The solid curve represents the best least squares fit
based on the 13 median values. Under the assumption that intrinsic spectral
properties at the sources could be described by simple power--law 
dependence, the average difference between jet and core spectral indices can 
be estimated as $\alpha_j-\alpha_c=-0.62\pm0.45$.

The three circles at the high redshift end of the plot in Fig.~\ref{SjSc} show
the upper limits of the jet to core flux density ratios for the most
distant ($z>4$) quasars imaged at 5~GHz with VLBI to date. The sources
\object{1251$-$407} ($z=4.46$, Shaver et al. \cite{SHA96}) and \object{1428+423}
($z=4.72$, Hook \& McMahon \cite{H&M98}) are represented by filled circles.
The open circle corresponds to the quasar \object{1508+572} ($z=4.30$, 
Hook et al. \cite{HOO95}) which also appeared to be unresolved,
however, at a considerably lower angular resolution ($\sim5$~mas) than
the other sources included in the sample (Frey et al. \cite{FS97}).

We note that in both cases available to date, radio structures in quasars
at $z>4$ (1251$-$407 and 1428+423) appear to be unresolved with a nominal resolution of $\sim$1~mas.
The third case, 1508+572, albeit with a lower resolution of 5~mas, does
not show a jet--like structure either.
Qualitatively, it is consistent with the overall trend that
steeper spectrum jets are fainter relative to flat spectrum cores at
higher redshift because the fixed 5~GHz observing frequency implies
high rest--frame frequency (for $z>4$  the emitted frequency
$\nu_{em}=\nu_{obs}(1+z)>25$~GHz). However, these sources
appear to be much more compact than expected from the general trend shown in
Fig.~\ref{SjSc}. Even in the neighboring high redshift bins ($3 < z < 4$), it is
unlikely that we find 3 randomly selected sources practically
unresolved.  A possible explanation for the observed compactness is
that the spectral indices of the jet components become steeper with 
frequency, which results in a relative fading of the components with
respect to the core at the high emitting frequencies ($\sim$25~GHz) 
of the largest redshift sources. 
Future multi--frequency VLBI observations of more $z>4$
radio loud quasars with the highest possible sensitivity and angular
resolution should answer the question whether these objects are indeed
intrinsically so compact or there is a strong observational selection effect
responsible for their particularly compact appearance.

\section{Conclusion}

We have presented 5~GHz VLBI images of ten extremely high redshift
($z>3$) quasars including the most distant radio loud quasar known to
date (1428+423, $z=4.72$). Most of the sources are well resolved and
their morphology is asymmetric. Based on fitted Gaussian source model
components, we have determined the jet to core flux density ratios. The
values obtained are typical of high redshift radio quasars for sources
in the redshift range $3 < z < 4$. However, the most distant radio loud
quasar, 1428+423, appears to be unusually compact.

\begin{acknowledgements}

We are grateful to the staff of the EVN, NRAO and Hartebeesthoek
observatories, and the MPIfR correlator for their support of
our project. We thank Joan Wrobel for assistance in preparation and analysis
of the global VLBI experiment of 1992 described in the paper,
and the referee for a number of very helpful suggestions.
ZP and SF acknowledge financial support received from the
European Union under contract CHGECT~920011, 
Netherlands Organization for Scientific Research (NWO) and the
Hungarian Space Office, and hospitality of JIVE and NFRA during their
fellowship in Dwingeloo. LIG acknowledges partial support from the EU
under contract no.  CHGECT~920011, the NWO programme on the Early
Universe and the hospitality of the F\"{O}MI Satellite Geodetic
Observatory, Hungary (supported in part through the contract No.
ERBCIPDCT940087 and by the Hungarian Space Office). LIG and RGM
acknowledge partial support from the TMR Programme, Research Network
Contract ERBFMRXCT~96--0034 ``CERES''. The National Radio Astronomy
Observatory is operated by Associated Universities, Inc. under a
Cooperative Agreement with the National Science Foundation. This
research has made use of the NASA/IPAC Extragalactic Data Base (NED)
which is operated by the Jet Propulsion Laboratory, California
Institute of Technology, under contract with the National Aeronautics
and Space Administration.

\end{acknowledgements}


\end{document}